\RequirePackage{fix-cm}

\documentclass[smallextended]{svjour3}

\usepackage{natbib}
\usepackage{balance}
\usepackage{fancyhdr}
\usepackage{amssymb}
\usepackage{amsmath}
\usepackage[figuresright]{rotating}
\usepackage{wrapfig}
\usepackage{multirow}
\usepackage{graphicx}
\usepackage{algorithm}
\usepackage{algorithmic}
\usepackage{url}
\usepackage{booktabs}
\usepackage{color}
\usepackage{moreverb}
\usepackage{balance}
\usepackage{amssymb,amsmath}
\usepackage{wrapfig}
\usepackage{multirow}
\usepackage{graphicx}
\usepackage{algorithm}
\usepackage{algorithmic}
\usepackage{epstopdf}
\usepackage{tabularx}
\usepackage{bigstrut}
\usepackage{url}
\usepackage{enumitem}
\usepackage{miniplot}
\usepackage{ifthen}

\usepackage{pifont}
\usepackage{bbding}
\usepackage{booktabs}
\usepackage{threeparttable}
\usepackage{array}
\usepackage{ragged2e}
\usepackage{etoolbox}
\usepackage{hyperref}

\makeatletter

\patchcmd{\NAT@citex}
  {\@citea\NAT@hyper@{%
     \NAT@nmfmt{\NAT@nm}%
     \hyper@natlinkbreak{\NAT@aysep\NAT@spacechar}{\@citeb\@extra@b@citeb}%
     \NAT@date}}
  {\@citea\NAT@nmfmt{\NAT@nm}%
   \NAT@aysep\NAT@spacechar\NAT@hyper@{\NAT@date}}{}{}

\patchcmd{\NAT@citex}
  {\@citea\NAT@hyper@{%
     \NAT@nmfmt{\NAT@nm}%
     \hyper@natlinkbreak{\NAT@spacechar\NAT@@open\if*#1*\else#1\NAT@spacechar\fi}%
       {\@citeb\@extra@b@citeb}%
     \NAT@date}}
  {\@citea\NAT@nmfmt{\NAT@nm}%
   \NAT@spacechar\NAT@@open\if*#1*\else#1\NAT@spacechar\fi\NAT@hyper@{\NAT@date}}
  {}{}

\makeatother

\newboolean{showcomments}
\setboolean{showcomments}{true}
\ifthenelse{\boolean{showcomments}}
{ \newcommand{\mynote}[2]{
    \fbox{\bfseries\sffamily\scriptsize#1}
    {\small$\blacktriangleright$\textsf{\emph{#2}}$\blacktriangleleft$}}}
{ \newcommand{\mynote}[2]{}}

\ifthenelse{\boolean{showcomments}}
{
  
}{
  \newcommand{\mynote}[2]{}
}

\begin{document}

\title{An overview of Web3.0 Technology: Infrastructure, Applications, and Popularity}


  \author{Renke Huang   \and Jiachi Chen \thanks{Jiachi Chen is the corresponding author.}
  	\and Yanlin Wang \and
	Tingting Bi \and Zibin Zheng}


\institute{Renke Huang \at
	School of Software Engineering, Sun Yat-Sen University, China\\
	\email{huangrk9@mail2.sysu.edu.cn}
	\and
	Jiachi Chen \at
	School of Software Engineering, Sun Yat-Sen University, China \\
	\email{chenjch86@mail.sysu.edu.cn}
	\and
	Yanlin Wang \at
	School of Software Engineering, Sun Yat-Sen University, China\\
	\email{wangylin36@mail.sysu.edu.cn}
	\and
	Tingting Bi \at
	Data61, Australia \\
	\email{tingting.bi@data61.csiro.au}
	\and
	Zibin Zheng \at
	School of Software Engineering, Sun Yat-Sen University, China\\
	\email{zhzibin@mail.sysu.edu.cn}}

\date{Received: date / Accepted: date}
\maketitle
\maketitle
\thispagestyle{fancy}
\lhead{}
\chead{}
\rhead{}
\lfoot{}
\cfoot{}
\cfoot{\thepage}
\renewcommand{\headrulewidth}{0pt}
\renewcommand{\footrulewidth}{0pt}
\pagestyle{fancy}
\cfoot{\thepage}

  \begin{abstract}
			Web3, the next generation of the Internet, represents a decentralized and democratized web. Although it has garnered significant public interest and found numerous real-world applications, there is a limited understanding of people's perceptions and experiences with Web3. In this study, we conducted an empirical study to investigate the categories of Web3 application and their popularity, as well as the potential challenges and opportunities within this emerging landscape. Our research was carried out in two phases. In the first phase, we analyzed 200 popular Web3 projects associated with 10 leading Web3 venture capital firms. In the second phase, we collected and examined code-related data from GitHub and market-related data from blockchain browsers (e.g., Etherscan) for these projects. Our analysis revealed that the Web3 ecosystem can be categorized into two groups, i.e., Web3 infrastructure and Web3 applications, with each consisting of several subcategories or subdomains. We also gained insights into the popularity of these Web3 projects at both the code and market levels and pointed out the challenges in the Web3 ecosystem at the system, developer, and user levels, as well as the opportunities it presents. Our findings contribute to a better understanding of Web3 for researchers and developers, promoting further exploration and advancement in this innovative field.
			
\end{abstract}


\section{Introduction}
\label{sec:introduction}
Web3, also known as Web 3.0, represents the third generation of the World Wide Web, which is usually built upon a decentralized blockchain network~\citep{sheridan2022web3}. Within this network, third parties cannot modify user data, and users maintain control over their data through public and private keys. Data changes adhere to predetermined agreements between users and the blockchain, establishing blockchain technology as the fundamental basis for Web3. Thus, Web3 is also known as the "read-write-own" web, as it empowers users to retain ownership of their data.

Web3 technology offers several key advantages. For example, the \textit{Immutability} feature ensures the data integrity of the Web3 project, and the \textit{Trustless} allows the Web3 projects do not rely on a trusted third party. Web3 also provides the feature to preserve user privacy and (\textit{Anonymity}) empowers users can control their data (\textit{Ownability}). In general, Web3 technology fosters a readable, writable, and ownable ecosystem, enabling users to manage and maintain control over their data through private and public keys. Due to the substantial potential of Web3 technology, Web3 applications have become increasingly prevalent in recent times. As of January 2023, over 13,091 Web3 applications have been deployed on various platforms~\citep{dappradar}. 

Although Web3 has garnered increased public interest, a clear definition remains elusive. Analyzing and understanding popular Web3 projects could yield insights into Web3 and its ecosystem. In this paper, we endeavor to address two research questions: \textbf{RQ1:} What are the categories of Web3 applications? \textbf{RQ2:} What is the popularity of Web3 projects?

To answer RQ1, we initially gathered 626 Web3 projects and selected the top 200 based on their investment amounts. We then manually analyzed the features and services provided by these projects, and conducted an open card sorting to cluster them into several groups. Finally, we discovered that the Web3 ecosystem comprises two categories, namely \textit{Web3 infrastructures} and \textit{Web3 applications}, with each further subdivided into several subcategories. Web3 infrastructure subcategories include DeFi, gaming, layer 2 scaling solutions, privacy, developer tools/service\-s, cross-chain interoperability, public chains, decentralized storage, and oracles. Web3 application subcategories encompass DeFi, NFT, Metaverse, DAO, and Web3 for traditional scenarios.

To answer RQ2, we collected code-related and market-related data for the 200 Web3 projects from GitHub and blockchain browsers (i.e., Etherscan, Bscscan, and Polygonscan). The analysis results
can assist Web3 practitioners and researchers in understanding the key attributes of Web3 projects, including dimensions such as code, market, and participant activities.


In addition, we also introduced the challenges and opportunities present within the Web3 ecosystem. Challenges were discussed from system, developer, and user perspectives. 
Challenges at system-level are interoperability, scalability, privacy; At developer-level are code security, incentive mechanism; and at user-level are usability and data recovery.
In terms of opportunities, Web3 features ensure security and grant users control over their data. Additionally, Web3 paves the way for new business models, enables novel forms of collaboration, and enhances transparency across industries.

The main contributions of this work are:
\begin{itemize}
   \item We conducted a comprehensive empirical study by analyzing 200 popular Web3 projects, from which we derived an understanding of the Web3 ecosystem. 
   \item  We uncovered the popularity of Web3 projects from code and market perspectives by analyzing code and market related data of popular Web3 projects. 
   \item We identified the challenges and opportunities present in Web3. This identification serves to inform researchers and developers of the critical issues and potential prospects within Web3.   
    \item We are open-sourcing a Web3 dataset~\citep{dataset}, which includes 200 popular Web3 projects that received funding from VC firms, as well as their corresponding code and market-related data.
 \end{itemize}

The remainder of the paper is organized as follows. Section~\ref{sec:rq1} and section~\ref{sec:rq2} show the answers for RQ1 and RQ2, respectively. Section~\ref{sec:cando} discusses the challenges and opportunities faced in Web3. Section~\ref{sec:related} discusses related work, and we conclude our study in Section~\ref{sec:conclusion}.

\section{Background}
\label{sec:background}

\subsection{Web1, Web2 and Web3}
Web1~\citep{noauthor_we1b_nodate-1} is a readable-only network, whereas Web2~\citep{noauthor_web2_2022} is a readable and writable network, and Web3~\citep{noauthor_web3_2022} is a readable and writable network that empowers users with data ownership. In Web1, the network primarily furnishes users with a collection of static pages for viewing purposes, devoid of interactive capabilities. While Web2 provides an array of dynamic pages, enabling user engagement with the network through activities such as content publication, file sharing, and payment. Nevertheless, Web2 is plagued by issues concerning the collection and storage of users' personal data on third-party platforms, potentially leading to unauthorized sales or illicit usage without the users' knowledge or consent. To address these challenges, Web3 has been proposed, which not only supports readable and writable but also ensures users maintain ownership of their data.

\subsection{Blockchain, Smart contract and Web3}
A blockchain consists of a sequence of records, referred to as blocks, which are interconnected and secured using cryptographic techniques. Each block is characterized by its transaction data, timestamp, and the hash value of the preceding block. The blockchain functions as a public ledger, with individual blocks housing records of diverse transactions. Rather than being stored in a centralized location, the blockchain is distributed across a network of nodes, each maintaining a copy. As a result, records are public and easily verifiable by all nodes, rendering data modification in the blockchain exceedingly costly. Modifying a block's transactions proves exceptionally challenging without attaining consensus among all nodes. Certain blockchains, such as Ethereum, provide generic computational capabilities via smart contracts. An Ethereum smart contract represents an account governed by an immutable program (i.e., bytecode). Users can initiate the execution of the bytecode by transmitting a transaction containing the specified execution parameters to the smart contract account. 

The blockchain and smart contract serve as the foundation for Web3, facilitating the running of applications on a decentralized network. Blockchain technology offers several key advantages for Web3, including: 1) \textbf{Immutability}. All the data stored on the blockchain cannot be changed. 2) \textbf{Trustless}. The interactions between users in a blockchain do not depend on a trusted third party. 3) \textbf{Availability}. The decentralization of blockchain provides high availability, as the failure of a single point on the network will not affect the normal running of a blockchain. 4) \textbf{Anonymity}. Users do not need to provide their real information, and their activities are usually anonymous in a blockchain.  5) \textbf{Ownability}. Blockchain allows users to control their data through private and public keys.
\section{RQ1: The Categories Of Web3 Applications}
\label{sec:rq1}
Web3, as a novel concept, currently still lacks a clear definition and deep understanding for the ecosystem. In this section, we introduce the details of how to find out the categories of Web3 applications. The results can help identify popular Web3 projects and provide further insights into the Web3 ecosystem.


\subsection{Approach} In order to gain an understanding of the Web3 ecosystem, we first need to find representative Web3 projects. Currently, the industry has already begun to explore the Web3 field for a long time and invested in various Web3 projects. Thus, we collected 200 popular Web3 projects invested by Web3 VC firms. Then, we conducted an analysis of the selected 200 Web3 projects, reviewing their documentation in order to provide a summary for each project. Finally, we performed open card sorting on these projects. The following provides further details on our approach.

\textbf{Step 1: Web3 projects collection.} In this step, we first identified 200 Web3 projects by examining the portfolios of Web3 VC firms. We began by using the "10 VC Firms Investing in Web3 Companies" list published by visible.vc\footnote{https://visible.vc/blog/web3-investors/} to identify 10 Web3 VC firms. 
We then manually examined all portfolios of these VC firms on Crunchbase\footnote{Crunchbase is a company providing business information about private and public companies. The website is: https://www.crunchbase.com} from January 2020 to June 2022, resulting in a total of 871 projects. For each project, we collected their name, official website URL, and financing amount received. Next, we used a keyword filtering method to identify Web3 projects. Specifically, we checked the official website of each project for the presence of the keyword ``Web3" or other blockchain-related keywords, such as ``blockchain", ``Web3 App", or ``decentralized"~\citep{wang2022exploring}.  We totally identified 626 (71.87\%) as Web3 projects. Finally, we selected the top 200 Web3 projects based on their investment amount from Web3 VC firms.

\textbf{Step 2: Documents analysis.} We conducted an analysis of the documentation for each Web3 project to understand their main functionalities. Specifically, the first and second authors conducted independent reviews of each project's documentation, identifying texts that could provide a preliminary summarization of the project's functionalities. For example, Uniswap~\citep{uniswap}, a decentralized exchange, the summarization text is ``a peer-to-peer system designed for exchanging ERC-20 Tokens". In cases where the documentation did not provide enough information about the project, we manually reviewed the other information, such as API instructions and use case examples, to extract relevant details regarding the project's functionality. Based on the information we extracted, we wrote a summarization for each project. If there is a disagreement on the summarization of a project between the authors, they discussed and consolidated their results.

\begin{figure}
    \centering
    \rotatebox{-90}{
        \includegraphics[width=1.1\linewidth]{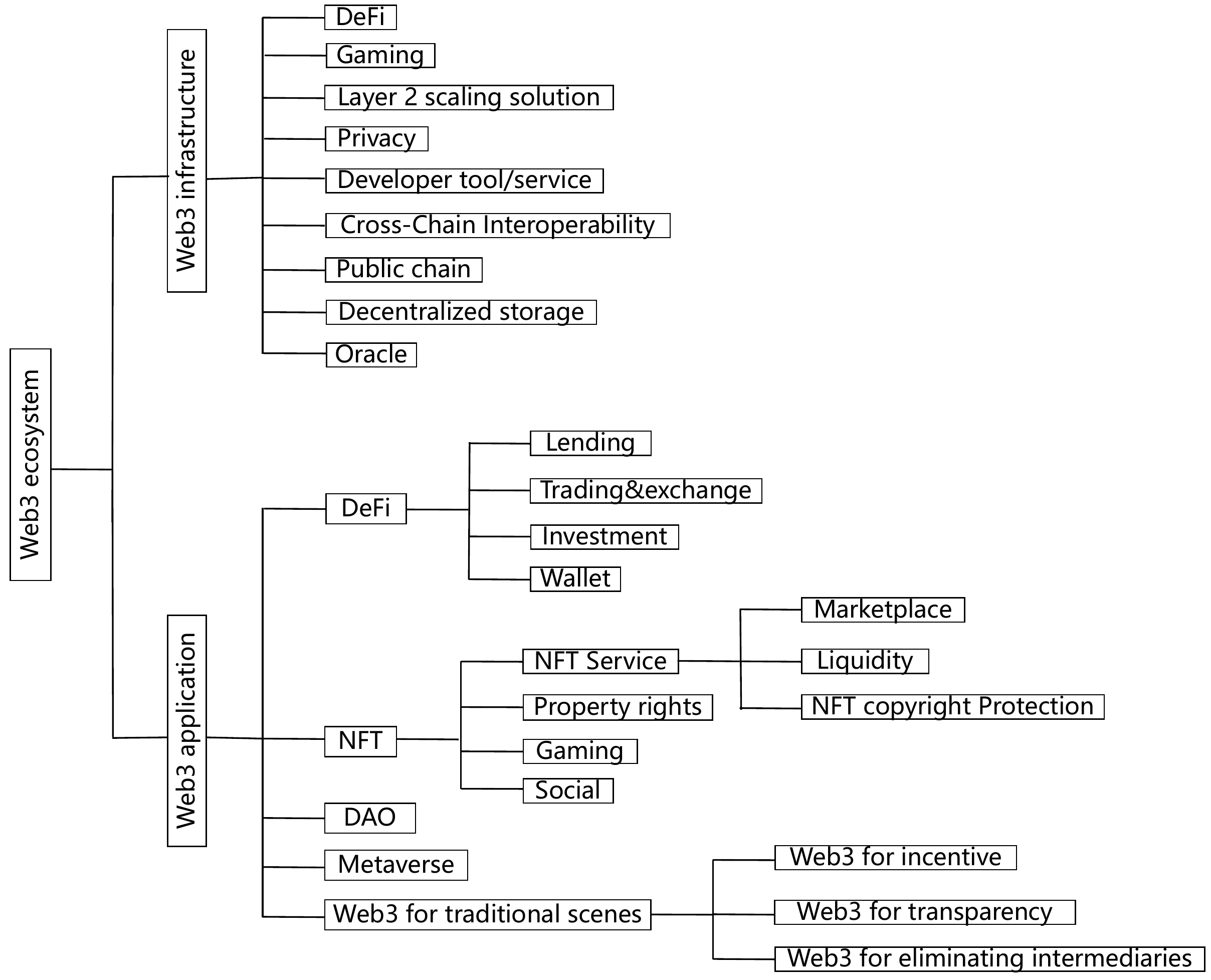}
    }
    \caption{The ecosystem of Web3}
    \label{Fig:web3ecosystem}
\end{figure}

\textbf{Step 3: Open card sorting.} After determining a summarization for each project, we performed card sorting\cite{spencer2009card} to identify categories based on these summarizations. Card sorting is a commonly used method for deriving categories from data. There are three main approaches to card sorting: closed, open, and hybrid. 
Given that the categories of Web3 projects are relatively unknown, we decided to follow an open card sorting approach to analyze these projects. During the open card sorting process, we created cards for each project that contained its name and summarization. Cards with similar summarizations were grouped together to form meaningful groups, each with a specific topic. These groups are equivalent to low-level subcategories, further evolved into high-level categories. The resulting hierarchical structure provided us with a clear categorization of the Web3 ecosystem. The first and second authors participated in the card sorting process and analyzed and verified each card. Ultimately, we identified two high-level categories within the Web3 ecosystem: Web3 Infrastructure and Web3 applications. Each category contains several subcategories (as shown in Figure~\ref{Fig:web3ecosystem}).

Figure~\ref{Fig:web3ecosystem} displays the results of the card sorting, providing an overview of the Web3 ecosystem. In the following, we provide an introduction for each of the two high-level categories, \textit{Web3 Infrastructure} and \textit{Web3 Applications}, and their subcategories. For each subcategory, we also highlight the Web3 project that received the highest financing, showcasing state-of-the-art projects within each subcategory.

\subsection{Web3 infrastructure}
Web3 infrastructure encompasses a variety of applications and solutions designed to support developers and enhance blockchain networks. Our analysis has identified nine subcategories of Web3 infrastructure: DeFi infrastructure, Gaming infrastructure, Layer 2 scaling solutions, Privacy infrastructure, Developer tools/services, Cross-chain interoperability services, Public chains, Decentralized storage, and Oracle services. In the following, we introduce each of these subcategories in more detail.

\subsubsection{DeFi infrastructure}
DeFi infrastructure is designed to simplify the development of token-swapping functionality by providing a standard set of APIs or SDKs. 

One of the prominent DeFi infrastructure projects is 0x~\citep{0x}, which has deployed smart contracts that support ERC20~\citep{erc-20} token swapping on Ethereum, as well as multiple EVM-compatible blockchains. These smart contracts aggregate liquidity from various decentralized exchanges (DEXs)~\citep{malamud2017decentralized} such as Uniswap~\citep{uniswap}. Then, the 0x protocol provides APIs for token-swapping functionality to developers. By utilizing these APIs, developers can easily integrate token-swapping functionality into their Web3 applications and share the liquidity from DEXs.

\subsubsection{Gaming infrastructure}
The gaming infrastructure facilitates the integration of in-game virtual assets into the blockchain for developers without blockchain experience.

Enjin~\citep{enjin} is a prominent gaming infrastructure that enables developers to design game using common programming languages such as C/C++, Java, and Python. Using the Enjin platform, developers can create in-game virtual assets, which are then assigned corresponding on-chain tokens. By utilizing the API provided by Enjin, developers can integrate these on-chain tokens into the game. Enjin is also responsible for monitoring requests from the game, processing requests on the blockchain, and returning data to the game.

\subsubsection{Layer 2 scaling solution}
Blockchain networks typically have low throughput, which limits the widespread adoption of Web3 applications. Layer 2 scaling solutions aim to increase network throughput without altering the underlying blockchain protocol. These solutions propose a framework for handling transactions off-chain and only reporting little information about the transaction on-chain, thereby achieving higher throughput.

Optimistic Rollups~\citep{schaffner2021scaling} is a representitive layer 2 scaling solution, which has been adopted by EVM-compatible layer 2 networks, e.g.,  Optimism~\citep{optimism}. This solution improves blockchain scalability~\citep{chauhan2018blockchain} by bundling multiple transactions on the layer 2 network, compressing them, and optimistically assuming their validity before submitting them to blockchain layer 1 for verification. During this process, any node on the layer 2 network can challenge transactions with fraud proofs when they observe invalid transactions. The node that submitted invalid transactions will be punished. This mechanism ensures the validity of the transactions submitted by the layer 2 network to a great extent.

\subsubsection{Privacy}
In most blockchain networks, user data is publicly visible, which poses a challenge to user privacy. Privacy infrastructure is designed to prevent users' transaction histories from being exposed to the public.

One of the representative privacy infrastructures is the Aztec protocol~\citep{aztec}. It consists of an Ethereum smart contract and a layer 2 network that adopts ZK-proofs~\citep{sun2021survey} technology. Users can transfer tokens from Ethereum to the Aztec layer 2 network by depositing tokens into the Aztec smart contract. In the layer 2 network, users can send tokens to others or interact with some Ethereum layer 1 smart contracts that are connected to the Aztec protocol. The transactions made by users in the Aztec layer 2 network are encoded as zkSNARK~\citep{petkus2019and} to protect users' transaction data from the public. Then, these transactions are bundled and sent to the Ethereum layer 1. Developers can integrate their smart contracts with the Aztec protocol using the API provided by the Aaztec protocol, enabling their Web3 applications to support user interaction in a private manner.

\subsubsection{Developer tool/service}
Developer tools \& services aim to assist developers in developing, testing, deploying, and managing smart contracts.

ConsenSys~\citep{consensys} is a popular project that provides development tools and services, including Infura~\citep{infura} and Truffle~\citep{truffle}. Infura is an Ethereum node provider that allows developers to connect Web3 applications to Ethereum without running their own nodes. It also provides APIs for Ethereum account management, smart contract deployment, transaction signing, and retrieving on-chain data. Truffle is an Ethereum-based Web3 applications development framework that provides smart contract compilation, testing, deployment, and interactive console features. It also provides libraries for Web3 application development.

\subsubsection{Cross-Chain Interoperability}
Communication between different blockchains was difficult. Cross-chain interoperability is a solution that aims to facilitate the flow of data and value across different blockchains.

Axelar~\citep{axelar} is a popular cross-chain interoperability solution that operates as an independent decentralized network with its validators. Axelar enables interoperability between blockchains by deploying smart contracts on these blockchains connected to Axelar. These smart contracts are controlled by a shared key that utilizes multi-party cryptography, which is divided into multiple parts and held by validators based on the amount of Axelar tokens they have staked. These validators run nodes on blockchains connected to Axelar to monitor on-chain activities. They are responsible for approving token cross-chain transfer requests from users after observing the deposits made by the users to the Axelar smart contract. When more than 50\% of the shares of the shared key from validators approve the request, the smart contract executes the token cross-chain transfer.

\subsubsection{Public chain}
The public blockchain is facing the challenge of the Blockchain trilemma~\citep{monte2020scaling}, which means it is difficult to meet the demands of decentralization, scalability, and security simultaneously.

Many public blockchains are exploring ways to maximize scalability while ensuring sufficient security. Near~\citep{near} is a representative project that proposes to improve scalability through sharding technology~\citep{wang2019sok} named Nightshade~\citep{skidanov2019nightshade}. In this sharding technology, data in a block is divided into multiple chunks based on the number of shards, with each shard responsible for one chunk. The shards are allocated to a portion of the validators in the network through a verifiable random function~\citep{micali1999verifiable}. The validator of a shard only needs to store and verify the chunk corresponding to the shard without storing and validating the complete block, thus improving scalability.

\subsubsection{Decentralized storage}
Storing data on popular blockchain networks like Ethereum can be expensive since the data needs to be stored on each node of the network. To address this issue, the decentralized storage network~\citep{benisi2020blockchain} was designed, which is essentially a blockchain. In this network, users' files are encrypted and stored on nodes of the network. The nodes only write encrypted metadata of the files (e.g., file location and file hash) into blocks of the network without the full files, thus reducing storage costs.

Filecoin~\citep{filecoin} is a popular decentralized storage network based on the IPFS protocol~\citep{benet2014ipfs}. In the Filecoin network, users pay fees based on file size and storage time, and then their files are stored on nodes. Nodes guarantee that the files have been stored and not deleted during storage by submitting Proof-of-Spacetime~\citep{benet2017proof} and Proof-of-Replication~\citep{benet2017proof} generated from the files to the Filecoin network.

\subsubsection{Oracle}
Oracle~\citep{beniiche2020study} is an interface used to deliver off-chain data to the blockchain for smart contracts to consume. However, delivering invalid or malicious data to the blockchain can potentially put smart contracts and user assets at risk.

To solve this problem, the Razor Network~\citep{razor} proposed a solution in which data providers are required to stake tokens. Providers of valid data will receive token rewards, while providers of invalid data will be fined. This mechanism enhances the reliability of the data provided to the blockchain, thus improving the security of smart contracts and user assets.

\subsection{Web3 applications}
Web3 applications refer to a set of applications that target users. In our study, we have identified five subcategories of Web3 applications: Decentralized Finance (DeFi), NFT, Metaverse, DAO, and Web3 for traditional scenes. Among them, the subcategories of DeFi, NFT, and Web3 for traditional scenes contain several smaller subcategories, which will be described in detail below.

\subsubsection{Decentralized Finance (DeFi)} 
The term ``DeFi" refers to the decentralized financial system that runs on top of the blockchain. In our dataset, the DeFi subcategory is further divided into four smaller subcategories: 1) lending, 2) trading \& Exchange, 3) investment, and 4) wallets. In the following, we will describe each of them in detail.

\textbf{1) Lending.} The lending protocol pools funds from multiple lenders to provide loans to borrowers and earns interest from the borrowers. The interest payments are then distributed to the lenders as returns on their lending.

Aave~\citep{aave} is a representative lending protocol that provides over-collateralized loans and flash loans~\citep{wang2020towards}. In an over-collateralized loan, the borrower is required to lock up crypto assets with a value exceeding the loan amount as collateral before the loan is released. In case the value of the collateral falls below a predefined threshold, the smart contract initiates an auction to recover the funds lent to the borrower and provide returns to the lender. Flash loans are uncollateralized loans available to developers with programming experience. In a flash loan, the borrower is only required to repay the entire loan amount and interest before the end of a single transaction without locking any assets.

\textbf{2) Trading \& Exchange.} Trading \& Exchange applications are a series of DEXs that enable on-chain token swapping without the need for a centralized custodian. However, implementing the order book model~\citep{abudy2018corporate} used by centralized exchanges (CEXs) on the blockchain often results in low efficiency.

In the trading and exchange field, Uniswap is a representative project that has adopted an Automated Market Maker (AMM)~\citep{xu2021sok} algorithm to replace the inefficient order book model. The AMM algorithm uses a liquidity pool mechanism for swapping, allowing users to have an instant token swap. The swap ratio is determined by the ratio of two tokens in the pool using a specific algorithm.

\textbf{3) Investment.} The investment protocol acts as a decentralized investment fund that encodes investment strategies into the smart contract. With this protocol, users can deposit funds into the smart contract, which then executes predefined rules to transact. After yielding profits, the profits are distributed to the users.

TokenSets~\citep{tokensets} is a popular protocol in the investment field that provides users with multiple portfolios. Each portfolio is managed by a smart contract, which transacts a specific set of tokens on DEXs based on predefined rules. TokenSets also allows investors to write their investment strategies into the smart contract and provide them to the public.

\textbf{4) Wallet.} Wallets are tools that allow users to manage their crypto assets and interact with Web3 applications. 

Metamask~\citep{metamask} is a popular wallet that enables users to create or recover their Ethereum accounts on any device by using human-readable phrases, without storing complex public and private keys. With Metamask, users can manage their ERC20 tokens and interact with Web3 applications.

\subsubsection{NFT}
NFT refers to a subcategory of Web3 applications that incorporate NFT elements~\citep{wang_non-fungible_2021}. This subcategory can be further divided into the following smaller subcategories: 1) NFT services, 2) property rights, 3) gaming, and 4) social. The details of each subcategory are described below.

\textbf{1) NFT Services.} 
NFT services consist of the following three small subcategories: \textbf{i)} NFT Marketplace, \textbf{ii)} liquidity protocol, and \textbf{iii)} NFT copyright protection service.

\textbf{i) NFT marketplace.} The NFT market is a platform for NFT trading. 

OpenSea~\citep{OpenSea} is currently the biggest NFT marketplace, offering users the ability to buy and sell various types of NFTs such as artworks, music, game assets, and more. Additionally, OpenSea allows users to mint NFTs without any prior experience in blockchain technology.

\textbf{ii) liquidity protocol.} NFTs are often associated with high prices and low liquidity. 

To address the liquidity issue, fractional.art~\citep{fractional} has proposed an NFT liquidity solution that involves dividing the ownership of an NFT. In this approach, the owner of the NFT locks NFT in fractional.art's smart contract. The smart contract then issues multiple ERC20 tokens based on the locked NFT, with each token representing proportional ownership of the NFT. This method effectively lowers the barriers to acquiring NFTs and increases their liquidity.

\textbf{iii) NFT copyright protection.} The NFT market is plag-ued by counterfeit NFTs, which are created by malicious users by uploading images of well-known NFTs and then minting NFTs to list on marketplaces.

To protect the copyright of NFTs, a project called Doppel~\citep{doppel} has provided a solution for detecting fake NFTs. This project uses computer vision and AI models to detect fake NFTs and report them. Currently, the company provides detection services for NFT on Ethereum, Solana, and Polygon.

\textbf{2) Property Rights.} NFT has been applied to protect digital/physical property rights. 

In terms of protecting digital property rights, a project called Royal.io~\citep{Royal} provides NFT-based music copyright protection. At Royal.io, artists can mint NFTs for their songs, with the song's name, hash value, and royalty share as the NFT's metadata. When the songs are purchased by third parties, the NFT holders will receive a portion of the royalties as their return. In terms of protecting physical asset property rights, a project called Origyn~\citep{origyn} mints high-definition images of real-world physical assets such as handbags, watches, and jewelry as NFTs, which serve as proof of property rights for these physical assets.

\textbf{3) Gaming.} NFT technology has also been adopted in the gaming industry, where in-game virtual assets are minted as NFTs, returning ownership of game data to the players. This type of game is also referred to as GameFi, enabling players to earn profits from the games.

Axie Infinity~\citep{noauthor_axie_nodate} is a prominent GameFi project where players can collect, breed, raise, and trade virtual pets, with each pet being stored as an NFT on the blockchain. Players can operate their pets to battle with other pets or sell their own pets to earn token rewards.

\textbf{4) Social.} NFT technology has also been applied in the social field, providing a way for users to express themselves freely without the control of tech giants. 

One notable project in this field is Mirror~\citep{mirror}, which is a decentralized blogging platform. On this platform, users can publish their content and mint them as NFTs. The content published by the user is then stored on IPFS, ensuring that the user's data cannot be arbitrarily altered or deleted. Furthermore, users can purchase NFTs minted by the creators to express their support.

\subsubsection{Metaverse}
Metaverse~\citep{gadekallu2022blockchain} refers to a virtual world built by computers that can interweave with the real world. 

Many metaverse projects have been built on top of the blockchain, characterized by the minting of characters or virtual assets of the virtual world as NFTs. One popular metaverse built for real estate is ``The Sandbox"~\citep{thesanbox}. In The Sandbox, users can purchase land on the game map and engage in secondary development. The in-game lands are recorded on the blockchain as NFTs, and users can also visit lands constructed by others.

\subsubsection{DAO}
DAO~\citep{mehar_understanding_2019} stands for decentralized autonomous organization, which operates through smart contracts. The smart contract manages the organization by executing the rules predefined by the organization's members.

In the DAO field, CreatorDAO~\citep{creatordao}is a typical example that brings together investors, creators, and supporters in a community. The DAO provides creators and their works with funding and resources to support their creative endeavors. When creators make profits from their works, the profits are automatically distributed to community members by the smart contract.

\subsubsection{Web3 for Traditional Scenes}
Web3 technology has also been adopted by many traditional industries to address challenges related to incentives, collaboration, transparency, trust, and etc. Based on the functionalities of Web3 in traditional industries, three subcategories have been identified: 1) Web3 for incentives, 2) Web3 for transparency, 3) Web3 for eliminating intermediaries.

\textbf{1) Web3 for incentives.} Some traditional industries are leveraging the incentive mechanism of Web3 by building their businesses on top of decentralized networks.

For instance, Helium~\citep{helium} is a blockchain-based wireless network operator that sells devices for providing wireless network coverage and mining. Users can purchase these devices and deploy them to provide wireless networks, thereby earning tokens. This incentive mechanism reduces the cost for enterprises to build and run their businesses while allowing users to benefit from traditional business scenes.

\textbf{2) Web3 for transparency.} Web3 has also been adopted by traditional industries to address trust issues by recording key business information on the blockchain.

Green Labs~\citep{greenlabs} and Blocery~\citep{blocery} are two companies that apply blockchain technology to the field of traceability. They record key information about products, such as production time, transfer history, and sales records, on the blockchain. The information recorded on the blockchain is publicly visible and tamper-proof, which enables the entire process of production and sales of products to be fully traceable, thereby increasing transparency and trust.

\textbf{3)Web3 for eliminating intermediaries.} Web3 eliminates the need for intermediaries in traditional business scenes by designing smart contracts to run the business.

For example, Dtravel~\citep{dtravel} is a decentralized travel platform on Ethereum. In this platform, travelers can directly book accommodations from property owners by interacting with smart contracts without having to go through online travel agents.

\section{RQ2: Popularity Of Web3 Projects}
\label{sec:rq2}

In this research question, we conducted an analysis of code-related and market-related data of Web3 projects to gain insights into their popularity. The findings of this research can assist Web3 practitioners and researchers in understanding the key attributes of Web3 projects, including dimensions such as code, market, and participant activities.

\subsection{Approach} To get insight into the popularity of Web3 projects on the code level and market level. We adopt the following methods to collect and analyze Web3 data.

\textbf{Step 1: Web3 data collection.} In this part, we collected code-related data of Web3 projects through Github and mark-et-related data of the Web3 projects through web3 blockchain browsers, i.e., Etherscan, Bscscan, and Polygonscan. In terms of code-related data, we first manually checked if the projects had open-source repositories on GitHub. After manually reviewing the 200 Web3 projects, we identified 96 open-source Web3 projects. Finally, we collected GitHub data for all 96 identified projects. In terms of market-related data, we obtained transaction and market cap data for each Web3 project between January 2022 and January 2023 by utilizing the APIs of blockchain browsers. 

\textbf{Step 2: Analysis of code level popularity.} For each subcategory of the Web3 infrastructure and Web3 applications, we calculated the number of open-source projects, as well as the total code lines and commits of code repositories for all projects(shown in table~\ref{Table:githubdata}), which were used to quantify the popularity of a particular subcategory at the code level. The subcategory has more projects, code lines, and commits, and the more popular it is. Additionally, we analyzed the blockchain platforms that Web3 applications were deployed on. Then we summarized the number of Web3 projects deployed on each blockchain platform to understand the popularity of blockchain platforms.

\textbf{Step 3: Analysis of market level popularity.} We divided Web3 projects into two groups: projects deployed on multiple blockchain platforms and projects deployed on a single blockchain platform. Each group includes the names of the Web3 projects, as well as their user numbers and market values. We analyze the median user numbers and market values of the Web3 applications in each group. By comparing the medians and maximum values of the two groups, we understand the impact of deploying Web3 applications on multiple blockchains versus a single blockchain in terms of user numbers and market value.
Following the aforementioned methodology, we divided Web3 projects into the open-source group and the non-open-source group. Then, we obtained the medians and maximum values of users and market value for the two groups, which reflect the impact of open-sourcing of Web3 projects in terms of user numbers and market value.
Additionally, for each subcategory of Web3 projects, we calculated the total number of users and market cap for all applications of each category(shown in table~\ref{Table:marketdata}), which were used to quantify the popularity of a particular subcategory at the market level. Finally, We analyzed the number of users of each Web3 application, figure~\ref{Fig:uaws} depicted the distribution of the number of users of each Web3 application.

In the following, we present the findings of the popularity of Web3 projects on two levels: code level and market level.

\begin{table}[!ht]
\centering
\caption{Code-related data of Web3 projects.}
\label{Table:githubdata}
\begin{tabularx}{\textwidth}{>{\raggedright\arraybackslash}X>{\centering\arraybackslash}X>{\raggedright\arraybackslash}X>{\raggedright\arraybackslash}X}
\hline
\textbf{Subcategories} & \textbf{Count} & \textbf{Code lines} & \textbf{Commits} \\
\hline
Public chains & 17 & 3,386,348 & 124,040 \\

\# Cross-Chain & 9 & 3,156,655 & 39,720  \\

Developer tool & 8 & 641,749 & 11,998  \\

\# Dstorage & 7 & 421,291 & 17,708  \\

Oracle & 3 & 364,499 & 872  \\

\# Gaming infra & 2 & 309,826 & 6,330  \\

Layer 2 scaling & 2 & 2,071,208 & 15,594  \\

Privacy & 2 & 436,352 & 7,960  \\

\# Defi infra& 1 & 105,496 & 17,137  \\

Lending & 14 & 595,999 & 11,205  \\

Trading\&exchange& 7 & 214,912 & 2,756   \\

Investment & 4 & 80,193 & 2,919  \\

DAO & 1 & 2,793 & 55   \\

Wallet & 2 & 162,163 & 4,533  \\

NFT Marketplaces & 2 & 117,305 & 2,977  \\

NFT Liquidity & 2 & 56,843 & 454  \\

\# NFT CP & 1 & 14,091 & 3 \\

\# Property rights & 2 & 105,070 & 806  \\

Gaming & 3 & 28,002 & 162   \\

Metaverse & 1 & 462,462 & 5,656  \\

Web3 for incentive & 2 & 98,049 & 3,804 \\

\mbox{\# Eliminating inters} & 4 & 20,992 & 321  \\
\hline
\end{tabularx}
\end{table}

\subsection{Code-level findings}
\begin{figure}
 \centering
		\includegraphics[width=0.7\textwidth]{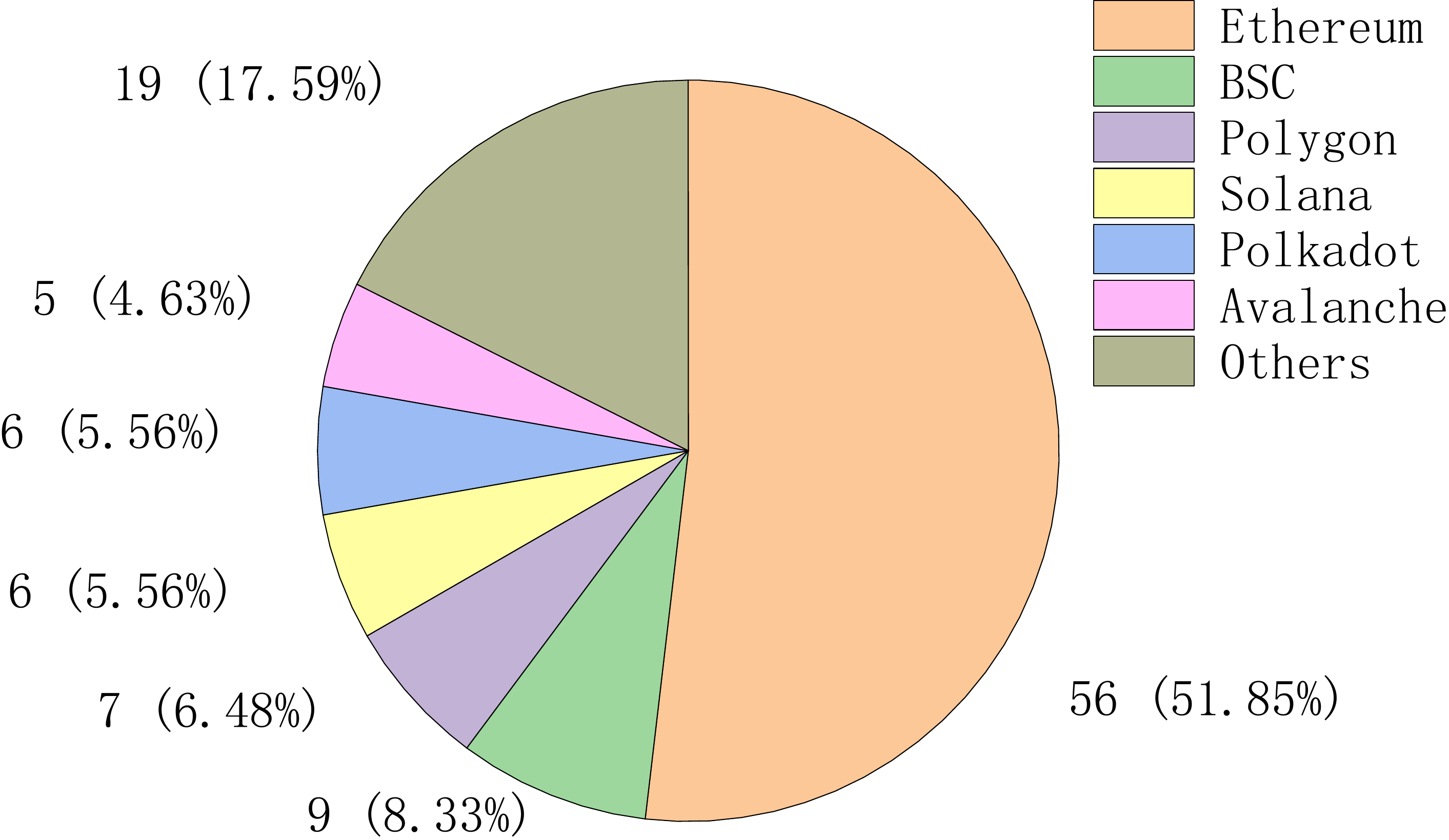} 
		\caption {Distributions of blockchain networks} 
		\label{Fig:platformdistribution}
\end{figure}

The code-level findings reflect the popularity associated with the developer. For example, the preferences for developing specific types of Web3 infrastructure and Web3 applications and the platform for deploying Web3 applications. Table~\ref{Table:githubdata} shows the code-related data of Web3 projects. The first column denotes the subcategories of open-source Web3 projects. In them, rows 1-9 denote the subcategories of Web3 infrastructure, while rows 10-21 denote the subcategories of Web3 applications. Among these subcategories, \textit{\# Cross-Chain} denotes the subcategory of Cross-chain interoperability.  \textit{\# Dstorage} denotes the subcategory of decentralized storage. \textit{\# Gaming infra} and \textit{\# Defi infra}denotes the subcategory of Gaming infrastructure and Defi infrastructure, respectively. \textit{\# NFT CP} denotes the subcategory of NFT copyright protection. \textit{\# Property rights} denotes the subcategory of NFT Property rights. \textit{\# Eliminating inters} denotes the subcategory of Web3 for eliminating intermediaries. Besides, Figure~\ref{Fig:platformdistribution} shows the distributions of blockchain networks. Following are some observations from the table.

\textbf{Observation 1: Public chain is the most popular subcategory in the Web3 infrastructure.} 
Table~\ref{Table:githubdata} shows that among open-source Web3 infrastructure projects, public chains are most numerous at 17, with the highest code lines (3,386,348) and commits (124,040).

\textbf{Observation 2: Lending is the most active subcategory in Web3 applications development.} As shown in table~\ref{Table:githubdata}, the Lending subcategory has the largest number of open-source projects among the Web3 application subcategories, with 12 projects. Moreover, the total number of code lines and commits are higher than other subcategories, at 595,999 and 11,205, respectively.

\textbf{Observation 3: Ethereum is the most popular blockc-hain network used to deploy Web3 applications.}
As shown in Figure~\ref{Fig:platformdistribution}, out of the 72 Web3 applications analyzed, 56 are deployed on the Ethereum network, accounting for 51.85\% of the total. 

\begin{figure}
 \centering
		\includegraphics[width=0.75\textwidth]{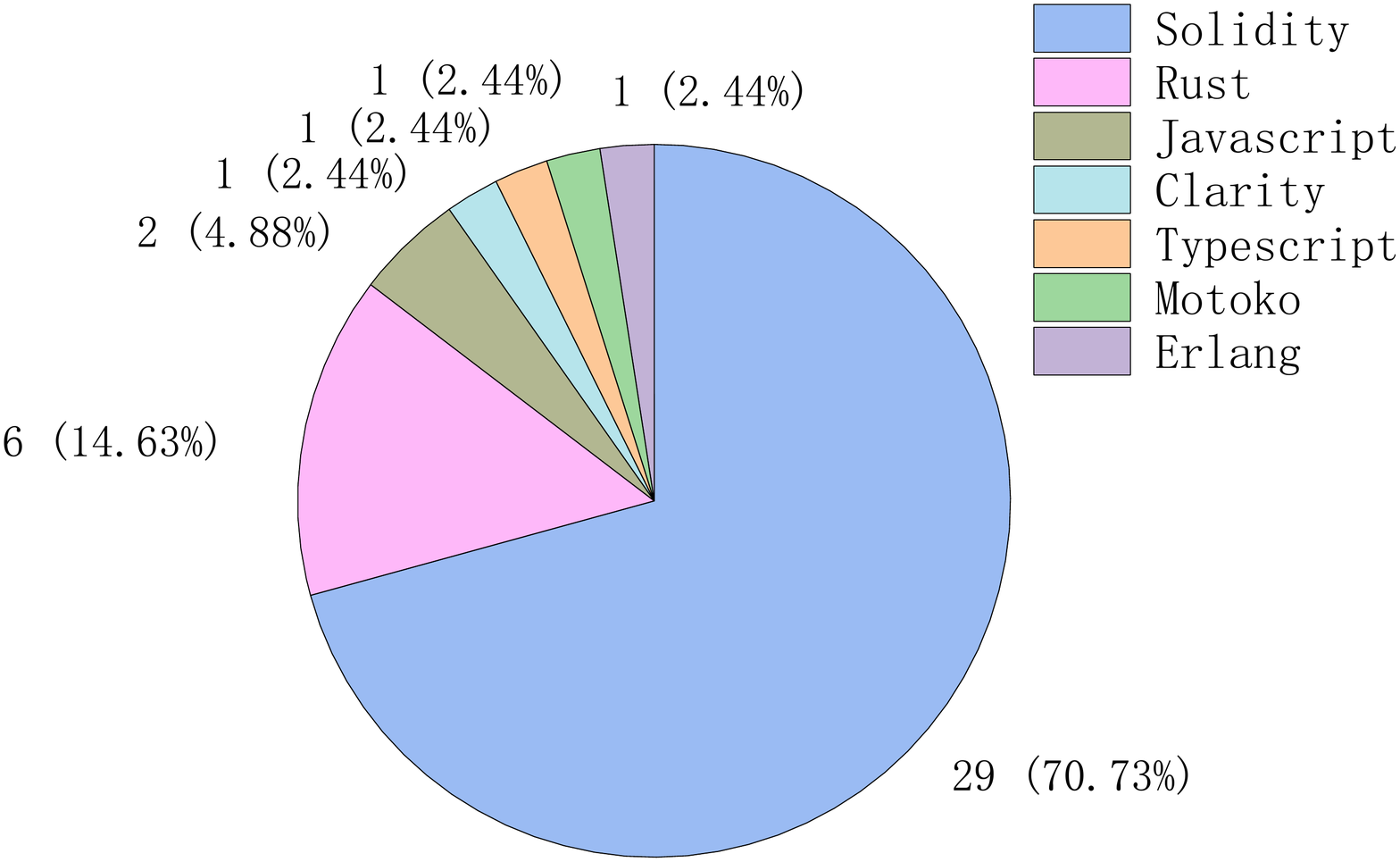} 
		\caption {Distributions of Web3 back-end programming languages} 
		\label{Fig:languages}
\end{figure}

\textbf{Observation 4: Solidity is the most popular programming language for Web3 applications' smart contract development.} Figure~\ref{Fig:languages} shows the distribution of back-end programming languages for 41 open-source Web3 applications. Out of the 41 open-source Web3 applications, 29 of them utilized Solidity as their programming language for smart contract(back-end) development.

\subsection{Market-level findings}
\begin{figure}
\centering
\includegraphics[width=0.8\textwidth]{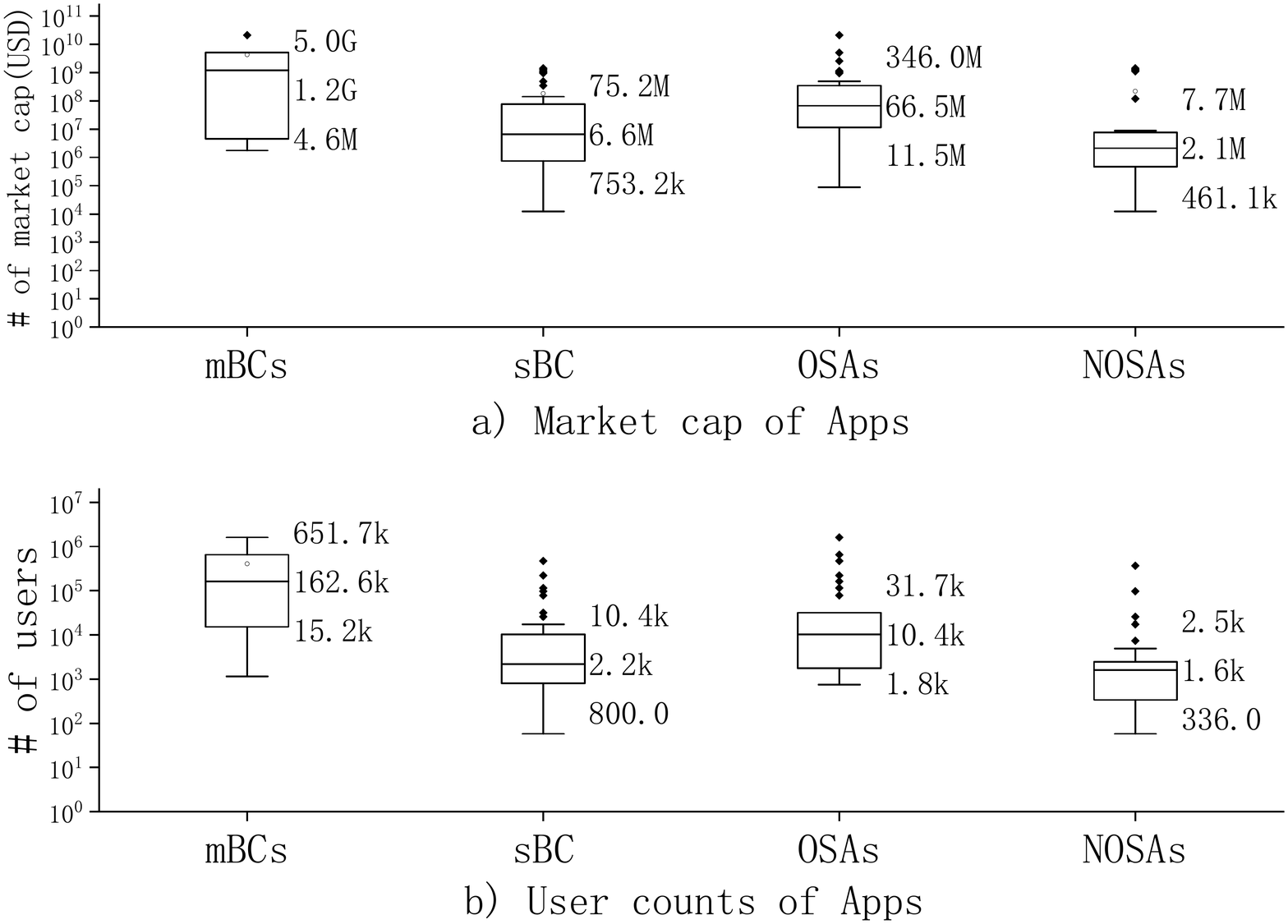} 
\caption{Distributions of user counts and market cap. \textbf{mBCs} denotes the Applications deployed across multiple blockchains. \textbf{sBC} denotes the applications deployed on a single blockchain. \textbf{OSAs} denote open source applications, while \textbf{NOSAs} denote non-open-source applications}
\label{Fig:2-1}
\end{figure}

\begin{figure}[!htbp]
\centering
\includegraphics[width=0.75\textwidth]{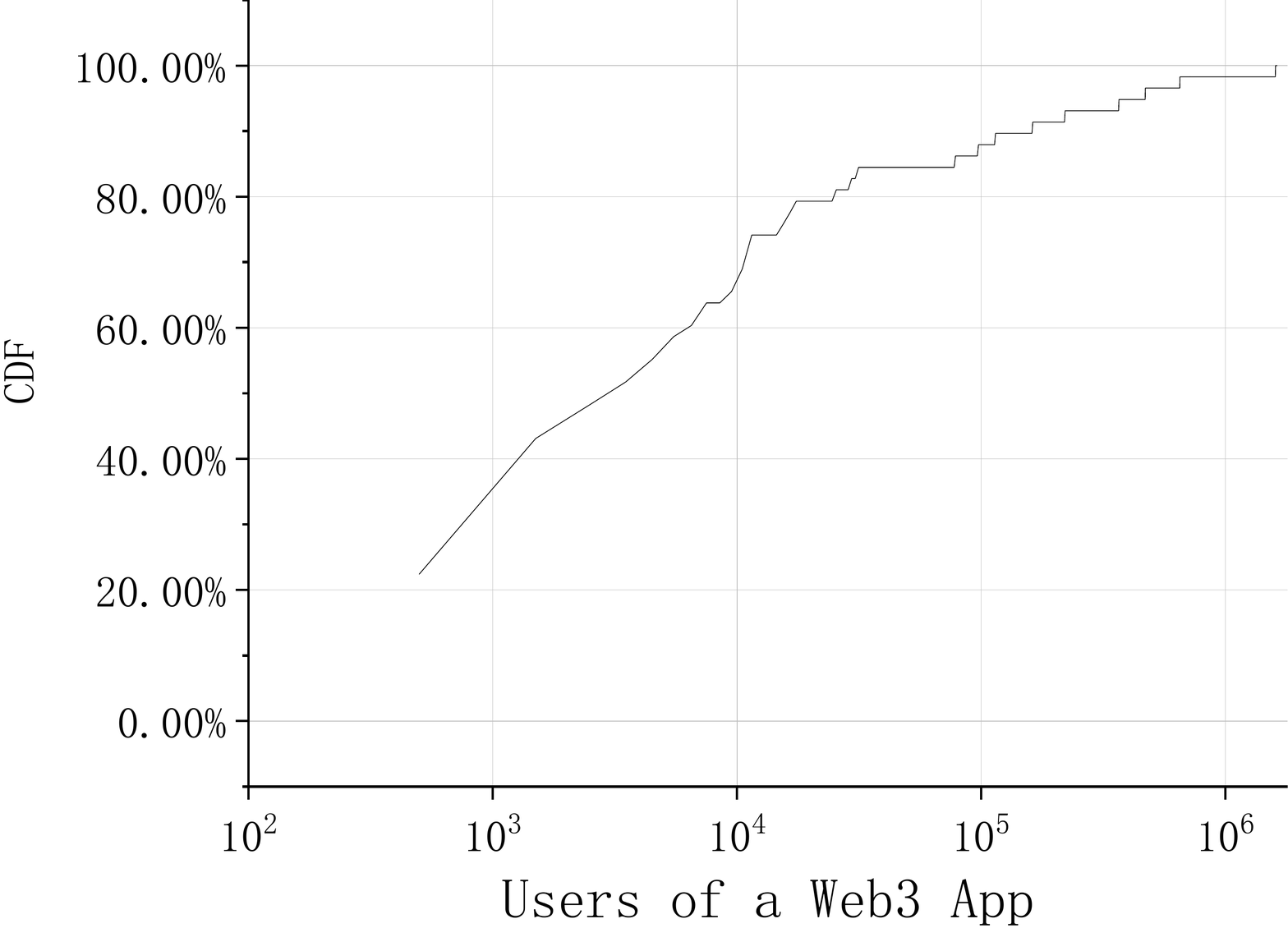} 
\caption {Number of users of a Web3 applications between January 2022 and January 2023.}\label{Fig:uaws}
\end{figure}

The findings at the market level reflect the popularity associated with user activity. These findings reflect the features of applications with a larger user base and market cap,  the most popular Web3 activities for users and the user count of most applications. Figure~\ref{Fig:2-1} shows the distributions of user counts and market cap, and Figure~\ref{Fig:uaws} illustrates the number of users of a Web3 applications between January 2022 and January 2023. Besides, Table~\ref{Table:marketdata} introduces market-related data of Web3 projects. Following are the observations from the market-level data. 

\textbf{Observation 5: Web3 applications deployed on multiple bloc-kchains have more users and a higher market cap.} In figure~\ref{Fig:2-1} b), the first and second boxes presents the distribution of the number of users for Web3 applications deployed on multiple blockchains and a single blockchain, respectively. It can be observed that the maximum value and median of the user numbers for Web3 applications deployed on multiple blockchains are higher than those deployed on a single blockchain. In figure~\ref{Fig:2-1} a), the first box and the second box represents the distribution of market capitalization for Web3 applications deployed across multiple blockchains and a single blockchain, respectively. It can be observed that the maximum and median of the market cap for Web3 applications deployed on multiple blockchains are higher than those deployed on a single blockchain.

\begin{table}[!htb]
\centering
\caption{Market-related data of Web3 projects. \textbf{\# Cnt} denotes the number of Web3 projects in the subcategory. \textbf{\# property rights} denotes the subcategory of NFT Property rights. \textbf{\# Eliminating inters} denotes the subcategory of Web3 for eliminating intermediaries.}
\begin{tabularx}{\textwidth}{@{\extracolsep{\fill}}>{\raggedright\arraybackslash}X>{\centering\arraybackslash}X>{\raggedright\arraybackslash}X>{\raggedright\arraybackslash}X} 
\hline
\textbf{Subcategories} & \textbf{\# Cnt} & \textbf{Users} & \textbf{Market cap} \\
\hline
Lending & 14 & 129,188 & \$1,827,265,578   \\

Trading\&exchange& 11 & 843,178 & \$5,831,446,783  \\

Investment & 2 & 22,287 & \$2,682,758,809 \\

NFT marketplace & 4 & 2,070,447 & \$24,687,589,059  \\

\# property rights & 4 & 16,422 & \$28,864,897  \\

Gaming & 17 & 193,249 & \$1,249,040,083  \\

Social & 3 & 18,780 & \$8,753,611  \\

Metaverse & 1 & 220,532 & \$950,498,005  \\

DAO & 4 & 49,375 & \$1,177,325,522   \\

Web3 for incentive & 3 & 477,043 & \$522,377,779  \\

\mbox{\# Eliminating inters} & 2 & 2,910 & \$16,049,029 \\ 

\hline
\label{Table:marketdata}
\end{tabularx}
\end{table}


\textbf{Observation 6: Open-sourcing Web3 applications can increase the number of users and market cap.} In figure~\ref{Fig:2-1} b), the third and forth box illustrate the distribution of the number of users of open-source and non-open-source Web3 applications, respectively. We can observe that the median of users for non-open source Web3 applications is lower, while open-source Web3 applications have a higher maximum number of users. In figure~\ref{Fig:2-1} a), the third and forth box display the distribution of the market cap of open-source and non-open-source Web3 applications, respectively. We can find that the maximum and median of market cap for open-source Web3 applications are higher than those for non-open-source Web3 applications.

\textbf{Observation 7: In Web3, NFT trading is the most popular activity for users.} As seen in Table~\ref{Table:marketdata}, the NFT market has the highest number of users at 2,070,447 and the highest market cap at \$24,687,589,059.

\textbf{Observation 8: About 65\% of popular Web3 applications have less than 10,000 users in a year.} The cumulative distribution function (CDF) of the number of unique users of the popular Web3 applications between January 2022 and January 2023 is shown in Figure~\ref{Fig:uaws}. 

\section{Challenges And Opportunities}
\label{sec:cando}

In this section, we point out the challenges faced by Web3 on three levels: system level, developer level, and user level. Then, we discuss the opportunities in Web3 ecosystem.

\subsection{Challenge in system level}
\label{C1}
As the underlying infrastructure that supports Web3 applications, problems in the blockchain system could potentially limit the development of Web3. We have identified five key aspects that may impact the Web3 ecosystem.

\textbf{S1) Scalability.} Blockchain's transaction throughput limits hinder Web3 applications from fulfilling real-time processing needs for millions of transactions. This problem also results in users paying high gas fees during network congestion. While some layer 2 scaling solutions have been proposed, they risk reduced decentralization and potential security issues.

\textbf{S2) Interoperability.} Different blockchains are not able to communicate with each other directly, which restricts the flow of data and tokens across different blockchains. Even though some cross-chain interoperability solutions have been proposed, they only support the transfer of tokens and not communication between applications on different blockchains.

\textbf{S3) Privacy.} The data stored on a blockchain is publicly visible, which makes user privacy more vulnerable to threats, especially when the relationship between a wallet and a physical identity is revealed. While some privacy protection solutions have been proposed, they are still unable to fully hide users' account balances and transaction histories.

\subsection{Challenge in developer level}
\label{C2}
In developing Web3 applications, the developers may face two challenges. The details are as follows:

\textbf{D1) Code security.} The code of Web3 applications deployed on the blockchain is publicly accessible, which makes it easier for hackers to exploit and launch attacks. Currently, there is a lack of mature code auditing tools, which puts pressure on developers to be cautious when writing application logic. Moreover, there have been instances where even audited code has been attacked, resulting in significant losses of crypto assets.

\textbf{D2) Incentive mechanisms design.} Poorly designed economic systems in Web3 projects can lead to failure. Many Web3 projects use token-based incentive mechanisms, but if the reward and consumption rules of the token are not well thought out, the token's value may plummet, causing the project to collapse. There have been several cases of Web3 projects failing due to issues with their economic systems.

\subsection{Challenge in user level}
\label{C3}
The following describes two challenges that users face while surfing on Web3.

\textbf{U1) Data recovery.} The loss of a private key can result in the loss of personal data and assets in Web3. Private keys serve as credentials for users to access and manage personal data in Web3. Losing them can lead to the crypto assets being locked and made unrecoverable. Therefore, it is necessary to design a mechanism for recovering user data.

\textbf{U2) Usability.} The user interface of Web3 applications is more complex and challenging to operate than that of Web2 applications. When interacting with a Web3 application, users are typically presented with a wallet interface containing complex strings, which can be confusing and result in potential asset loss. For example, a user may inadvertently sign a ``setapprovalforall" transaction launched by a Web3 application if they cannot understand the transaction details, resulting in the transfer of all their assets to another party. Currently, many phishing scams exploit this vulnerability, putting users' assets at risk.

\subsection{Opportunities in Web3}
\label{O1}
We present multiple opportunities that Web3 brings in the following:

\textbf{O1) Business model.} Web3 has motivated the emergence of new business models through its decentralized design and smart contracts. For example, in the finance industry, DeFi provides users with new ways to store, trade, and invest their assets. In the mobile communication industry, Helium~\citep{helium} explores the deployment of network infrastructure in a distributed manner, allowing people to participate and profit from network deployment and operations. These examples demonstrate how Web3 enables the creation of innovative business models through decentralization.

\textbf{O2) Collaboration.} DAOs facilitate new collaboration models. They transform the traditional organizational model through decentralized governance and transparent operations, providing a significant opportunity to create a more efficient and democratic form of organization. MakerDAO~\citep{MakerDAO} is a DeFi protocol built on Ethereum that allows users to obtain stablecoins by staking crypto tokens. In MakerDAO, important decisions such as adjustments to fees and the addition of staking token types are decided through a voting process by MakerDAO token holders.

\textbf{O3) Transparency} Web3 introduces a new method of information recording that enhances transparency in various business scenarios. Recording information on a blockchain enables it to be publicly visible and immutable, thereby addressing issues related to trust and transparency. For instance, supply chain management utilizes blockchain to record relevant information, resulting in increased transparency, trustworthiness, and management efficiency.

\textbf{O4) Ownership of data.} Web3 restores data ownership to users. In Web3, personal user data is stored on the blockchain and controlled by the user who holds the private key. This eliminates the control of centralized institutions over user data. For example, when a game adopts NFT technology, the game operator cannot modify the user's data, ensuring the security of user data.

\section{Related Work}
\label{sec:related}
There have been several previous studies on the topic of Web3, the details are as follows.

Wang et al.~\citep{wang2022exploring} conducted the first empirical investigation of existing Web3 projects, extracting a concise backbone model that delineates participating roles and operational workflows of Web3 projects. Subsequently, they proposed twelve distinct Web3 architectural designs, capturing the operational mechanisms of typical Web3-based applications. In their study, they deconstructed a complete Web3 service into three-tiered components, based on data workflow considerations. Data within each component can be managed through on-chain, off-chain, or hybrid methodologies. The identified design types effectively represent the full spectrum of potential Web3 applications and service combinations. To evaluate the merits of each architectural design, the authors meticulously assessed each configuration using a variety of property metrics, derived from classic blockchain systems. Additionally, they explored which participating entities stand to gain the most advantages under varying design types. Expanding the scope of their research beyond architectural design, they examined the broader implications of Web3, discussing its impacts, opportunities, and challenges. 

Sheridan et al.~\citep{sheridan2022web3} delineated the fundamental components of Web3 implementation, encompassing blockchain networks, Web3 programming languages, Web3 libraries, smart contracts, and wallets. Furthermore, they presented an overview of Web3 developers, as well as the impacts and risks associated with Web3. Lastly, they projected possible future integrations of Web3 with other emerging technologies.

Yu et al.~\citep{yu2022towards} delved into the essence of Web3 by examining an extensive range of real-world Web3 projects. They conducted a comparative analysis between existing Web2 solutions and self-proclaimed Web3 projects to identify the fundamental characteristics of Web3 applications and their dependencies, as well as to distinguish their differences from traditional Web2 applications. Through a thorough investigation, the authors proposed a seamless transition framework, dubbed WebttCom, aimed at facilitating the transition from Web2 to Web3. The innovative WebttCom framework offers efficient and reliable access control and user management across both decentralized Web3 and centralized Web2 environments. Furthermore, it presents a viable method for implementing the transition with widely-used Software as a Service platforms, and showcases a practical use case implemented using this framework. In order to assess the framework's effectiveness from the developers' perspective, they conducted interviews with five proficient developers. The feedback obtained from these interviews demonstrated that the research question was adequately addressed by the proposed WebttCom framework. Subsequently, the authors put forth several recommendations for enhancing the WebttCom framework, which includes extending its compatibility to incorporate additional blockchain platforms and exploring further potential business cases. 

Liu et al.~\citep{liu2021make} introduce the first comprehensive and quantifiable metric, referred to as verifiability, for delineating the Web3 era, which is grounded in empirical observations and rational analysis of the evolution of blockchain infrastructures in recent years. In light of this characterization, we identify three fundamental infrastructural enablers for Web3: individual blockchains with smart contract capabilities, centralized or federated state publishers, and interoperability platforms designed to bridge the gaps between these disparate systems. Subsequently, the authors offer an in-depth exploration of one of these three core enablers: HyperService, the pioneering interoperability platform that seamlessly connects heterogeneous blockchains and federated or centralized state publishers to establish a unified, cohesive computing platform for Web3 applications. The researchers implement a prototype of HyperService, consisting of approximately 62,000 lines of code, and assess its performance using three distinct categories of crosschain decentralized applications. Experimental results demonstrate that HyperService imposes an end-to-end dApp execution latency on the order of seconds, while also exhibiting horizontal scalability in the platform.

In contrast to previous studies, our research focuses on delineating the Web3 ecosystem through popular projects, analyzing overviews and representative applications for each field, and evaluating the popularity of Web3 projects. Additionally, we underscored the opportunities presented by Web3 and emphasized the challenges it poses for blockchain systems, developers, and users.

\section{Conclusion}
\label{sec:conclusion}

In this study, we conducted a comprehensive empirical study to gain insights into the Web3 ecosystem. We first selected the top 200 Web3 projects and performed open card sorting to investigate the categories of Web3 application, which comprises two main categories: Web3 infrastructure and Web3 applications, each containing several subcategories.  We introduced each subcategory and its representative project to provide a comprehensive overview. Then, we collected code-related and market-related data for the 200 Web3 projects from GitHub and blockchain browsers. This information helps developers better understand and navigate the Web3 landscape. In addition, we discussed the challenges facing the Web3 ecosystem at the system, developer, and user levels, while also highlighting the opportunities it presents. Our findings empower researchers, developers, and other stakeholders to deepen their understanding of and contribute more effectively to the evolving Web3 ecosystem.

\section{Declarations}
\label{sec:declarations}
\begin{flushleft}
Competing Interests \\
The authors declared that they have no conflict of interest. \\
\end{flushleft}

\begin{flushleft}
Data Availability \\
The datasets generated during and/or analyzed during the current study are available in the Github repository, \href{https://github.com/popularWeb3projects/Web3-dataset}{https://github.com/popularWeb3projects/Web3-dataset}.
\end{flushleft}

\balance
\bibliographystyle{spbasic}
\bibliography{ref}

\end{document}